\pdfoutput=1
\documentclass{aa}  

\usepackage{graphicx}
\usepackage{setspace} 
\usepackage[varg]{txfonts}
\usepackage{natbib}
\usepackage{textcomp, gensymb} 
\usepackage{gensymb}
\bibpunct{(}{)}{;}{a}{}{,}

\begin{document}\small

   \title{Solving the distance discrepancy for the open cluster NGC~2453 \thanks{Based on data collected at the 2.5m duPont telescope located at Las Campanas Observatory, Chile (program ID CHILE-2013A-157)}}
   \subtitle{The planetary nebula NGC~2452 is not a cluster member}

   \author{D. Gonz\'alez-D\'iaz\inst{1,}\,\inst{2}
          \and
          C. Moni Bidin\inst{1}
          \and
          E. Silva-Villa\inst{2}
          \and
          G. Carraro\inst{3}
          \and
          D. Majaess\inst{4,5}
          \and
          A. Moitinho\inst{6}
          \and
          C. Orquera-Rojas\inst{1}
          \and
          C. A. L. Morales Mar\'in\inst{1}
          \and
          E. Morales-Campa\~na\inst{1}
          }

   \institute{Instituto de Astronom\'ia, Universidad Cat\'olica del Norte, Av. Angamos 0610, Antofagasta, Chile,\\              			\and
           Instituto de F\'isica, Universidad de Antioquia (UdeA), Calle~70~52-21, Medell\'in, Colombia
             \\
                \email{danilo.gonzalez@udea.edu.co}
         \and
             Dipartimento di Fisica e Astronomia, Universit\'a di Padova, Vicolo Osservatorio 3 I-35122, Padova, Italy\\
         \and
             Mount Saint Vincent University, Halifax, Nova Scotia, Canada\\
         \and
Saint Mary's University, Halifax, Nova Scotia, Canada\\
         \and
             SIM - Faculdade de Ci\^{e}ncias da Universidade de Lisboa, Ed. C8, Campo Grande, 1749-016, Lisboa, Portugal\\
             }

   \date{Received Month DD, YYYY; accepted Month DD, YYYY}

 
  \abstract
  {The open cluster (OC) NGC~2453 is of particular importance since it has been considered to host the planetary nebula (PN) NGC~2452, however their distances and radial velocities are strongly contested.}
  {In order to obtain a complete picture of the fundamental parameters of the OC NGC~2453, 11 potential members were studied. The results allowed us to resolve the PN NGC~2452 membership debate.}
  {Radial velocities for the 11 stars in NGC~2453 and the PN were measured and matched with \textit{Gaia} data release 2 (DR2) to estimate the cluster distance. In addition, we used deep multi-band UBVRI photometry to get fundamental parameters of the cluster via isochrone fitting on the most likely cluster members, reducing inaccuracies due to field stars.}
  {The distance of the OC NGC\,2453 (4.7 $\pm$ 0.2 kpc) was obtained with an independent method solving the discrepancy reported in the literature. This result is in good agreement with an isochrone fitting of 40-50 Myr. On the other hand, the radial velocity of NGC\,2453 ($78 \pm 3$~km\,s$^{-1}$) disagrees with the velocity of NGC\,2452 ($62 \pm 2$~km\,s$^{-1}$). Our results show that the PN is a foreground object in the line of sight.}
  {Due to the discrepancies found in the parameters studied, we conclude that the PN NGC~2452 is not a member of the OC NGC~2453.} 
  

\maketitle

%
\section{Introduction}
\label{sec:Introduction}
  
The planetary~nebula~(PN)~/~open~cluster~(OC) pair NGC\,2452\,/\,NGC\,2453 has been widely studied, and the membership of the PN to the stellar cluster has been heavily contested. The measurements of both distance and age of the cluster ($\alpha_{2000}=07^{h}47^{m}36\fs7$, $\beta_{2000}=-27\degr 11^{\prime} 35^{\prime\prime}$) in the literature have not reached an agreement. The early photometric study and main sequence (MS) fit of 21 cluster stars by \citet[hereafter MF]{moffat1974}\defcitealias{moffat1974}{MF} established a distance of $d\sim 2.9$~kpc and an age of $\tau\sim 40$~Myr. Other studies approximately agreed, proposing cluster distances in the range $d$ $\approx$2.4-3.3~kpc \citep{glushkova1997,hasan2008}, while \citet{gathier1986} obtained almost twice the distance value (5.0$\pm$0.6~kpc) via Walraven photometry on five stars previously reported as members by \citetalias{moffat1974}. Later, \citet{mallik1995} revealed a deeper  MS of the cluster by means of $BVI$ photometry. These latter authors determined a distance of about $d\approx 5.9$~kpc, with a mean age of $\tau\approx 25$~Myr, but they also showed that the best fit depended on which stars were considered cluster members. In fact, the line of sight to the PN/OC pair is highly contaminated by field stars belonging to the Puppis OB associations and the Perseus arm \citep{peton1981,majaess2007}. This complex mix of different stellar populations in the color-magnitude~diagram (CMD) inevitably adds uncertainty to the results of an isochrone fit, which could be easily affected by field stars. 

NGC\,2452 ($\alpha_{2000}=07^{h}47^{m}26\fs26$, $\beta_{2000}=-27\degr 20^{\prime} 06\farcs83$) is a massive PN \citep{cazetta2000}, whose progenitor must have been an intermediate-mass MS star close to the upper limit allowed for PN formation. This is consistent with the $\sim$ 40 Myr age of NGC\,2453 proposed by \citetalias{moffat1974} and \citet{moitinho2006}, which implies a turnoff mass of $\approx$7~M$_{\odot}$. The cluster age is an important parameter to discard membership to young OCs because evolved stars in clusters younger than $\sim$30~Myr are thought to end as type-II supernovae rather than forming a PN \citep[see][hereafter MB14]{majaess2007,bidin2014}.\defcitealias{bidin2014}{MB14} Distance estimates for this PN can also be found in the literature, from 1.41 kpc, passing through 2.84 kpc, to 3.57 kpc \citep[respectively, among others]{khromov1979,stanghellini2008,gathier1986}. 
The value obtained by \citet{gathier1986} ($d=3.57\pm0.5$~kpc)
from a reddening-distances diagram was very different from the cluster value derived from zero-age MS (ZAMS) fitting in the CMD and two-color diagram (TCD, $\sim$5~kpc). However, their estimate of the PN reddening ($E_{B-V}=0.43\pm0.5$) roughly matched the literature value for the cluster, which is in the range $\sim$0.47-0.49 \citep{moffat1974,gathier1986,mallik1995}.

The association between NGC\,2453 and NGC\,2452 has been proposed and studied by many authors, in light of their angular proximity in the sky (angular separation $\sim 8\farcm5$) and the data available (see, e.g., \citetalias{moffat1974}, \citealt{gathier1986}, \citealt{mallik1995}, \citetalias{bidin2014}). Nevertheless, the results have not been conclusive. \citet[]{moffat1974} found coincidences between the radial velocity (RV) of the PN measured by \citet[][68~km~s$^{-1}$]{campbell1918} and that of an evolved blue giant star in the cluster (67$\pm$4~km~s$^{-1}$). Subsequent measurements yielded consistent RVs for the PN in the range $\sim$62-68~km~s$^{-1}$ \citep{meatheringham1988,wilson1953,durand1998}. Nevertheless, \citet{majaess2007} advocated additional observations needed to evaluate potential membership. \citetalias{bidin2014} recently studied the RV of ten stars in the cluster area, supporting the cluster membership of NGC\,2452. However, they claimed that their result was not definitive, because the identification of cluster stars was problematic.

In this work, we have adopted the methodology followed by \citetalias{bidin2014} and expanded the sample to 11 potential members to assess the membership of NGC\,2452 to NGC\,2453 via RV measurements on intermediate-resolution spectra. In addition, deep $UBVRI$ photometry was paired with data from\textit{ Gaia}'s second data release \citep[DR2,][]{gaia2018} to revise the cluster distance and to accurately determine its fundamental parameters.

%
\section{Observations and data reduction}
\label{s_data}

%
\subsection{Spectroscopic data}
\label{ss_dataspec}

The intermediate-resolution spectra of 11 bright stars of NGC\,2453 were collected on April 18, 2013, during one night of observations at the duPont 2.5m telescope, Las Campanas, Chile. The targets were selected on the IR CMD based on 2MASS data, prioritizing the brightest stars next to the cluster upper MS. The SIMBAD names and 2MASS photometry of the targets are given in Table~\ref{Tab:RV_Dis}. The 1200 line/mm grating of the B\&C spectrograph was used with a grating angle of 16$\fdg$67 and a 210$\mu$ slit width, to provide a resolution of 2~\AA\ (R=2200) in the wavelength range 3750-5000~\AA. Exposure times varied between 200 and 750s, according to the magnitude of the  target. A lamp frame for wavelength calibration was collected regularly every two science spectra during the night.

\begin{figure}
        \centering
                \includegraphics[trim = 19mm 0mm 0mm 10mm, clip, width=10cm]{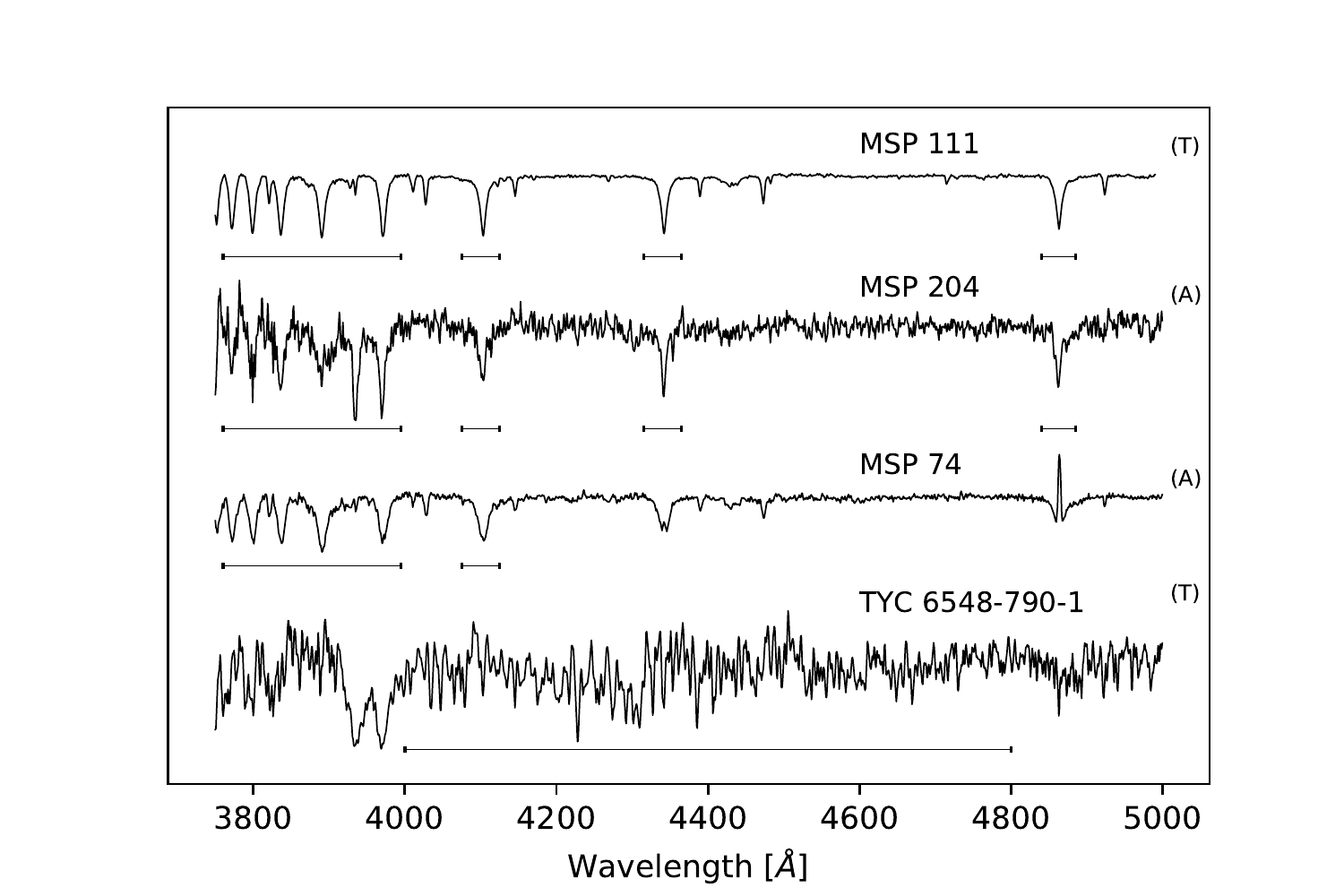}
        \caption{Examples of reduced spectra. The wavelength intervals used in RV measurements are shown as horizontal lines. The spectra are labeled as T and A for `target' and `additional stars', respectively.  The spectra have been shifted vertically to avoid overlap.}
        \label{Fig:spectra}
        \end{figure}

The spectra were reduced by means of standard IRAF routines. Figure~\ref{Fig:spectra} shows some examples of the final result.The resulting S/N for the selected targets was typically S/N=80–120. Non-target stars fell regularly in the slit in almost all exposures, because both the OC and the surrounding low-latitude Galactic field are very crowded. Their spectra were reduced and analyzed in the same way as those of our targets, but the resulting spectra were of much lower quality (S/N$\approx$10--30). We hereafter refer to ``target'' and ``additional'' stars, to distinguish between the selected objects and the stars that fell by chance in the spectrograph slit.

During the same run, we collected three spectra of the PN NGC\,2452. The first one was acquired centering the slit at the optical center of the nebula, where a bright spot was seen. The second and third spectra focused on the northern and southern regions, respectively. The reduction of these data proceeded as in the case of cluster stars, but the frames of a bright RV standard star were used during extraction to trace the curvature of the spectra on the CCD. The PN is an extended object, and its spectrum covered several pixels in the spatial direction. We performed both a narrow (8~pixels, $\sim 5\arcsec$) and a wide (20~pixels, $\sim 65\arcsec$) extraction for the northern and central spectra, but only a narrow extraction for the southern one because the flux was too faint outside $\pm4$~pixels from the center.

%

%
\subsection{Photometric data}
\label{ss_dataphot}

Our study is based on the optical $UBVRI$ photometric catalog presented by \citet{Moitinho01}. The data were acquired in January 1998 at the CTIO 0.9m telescope, with a $2048\times2048$ Tek CCD, with a resulting $0\farcs39$ pixel scale and a $13\arcmin\times13\arcmin$ useful field of view. The frames were processed with standard IRAF routines, and the shutter effects were corrected applying a dedicated mask prepared during the reduction. We refer to \citet{Moitinho01} for a very detailed presentation of observations and data reduction.

%
\subsection{\textit{Gaia} distances}
\label{sub:GaiaDistances}

Parallaxes and proper motions for program stars were obtained from the \textit{Gaia} DR2\footnote{Gaia Archive: https://gea.esac.esa.int/archive/} catalog. We added +0.029~mas to all \textit{Gaia} parallaxes, as advised by \citet{Lindegren18}, to account for the zero-point offset reported by \citet{Lindegren18} and \citet{Arenou18}. Following the guidelines of \citet{Luri18}, we employed a Bayesian method to infer distances from parallaxes through a model error and a priori assumption. Because the fractional errors on parallax are $f_\omega=\sigma / \varpi\leq$0.24 most program stars, we used the exponentially decreasing space density function (EDSD) as a prior, as described by \citet{BailerJones15}. A complete Bayesian analysis tutorial is available as Python and R notebooks and source code from the tutorial section on the \textit{Gaia} archive\footnote{https://github.com/agabrown/astrometry-inference-tutorials/}. Proper motions and distances computed from \textit{Gaia} DR2 parallaxes are shown in Table \ref{Tab:RV_Dis}. Upper and lower indices correspond to maximum and minimum distances in the error interval, respectively.

%
        \begin{figure}
        \centering
                \includegraphics[trim = 5mm 0mm 0mm 10mm, clip,angle=0,width=10cm]{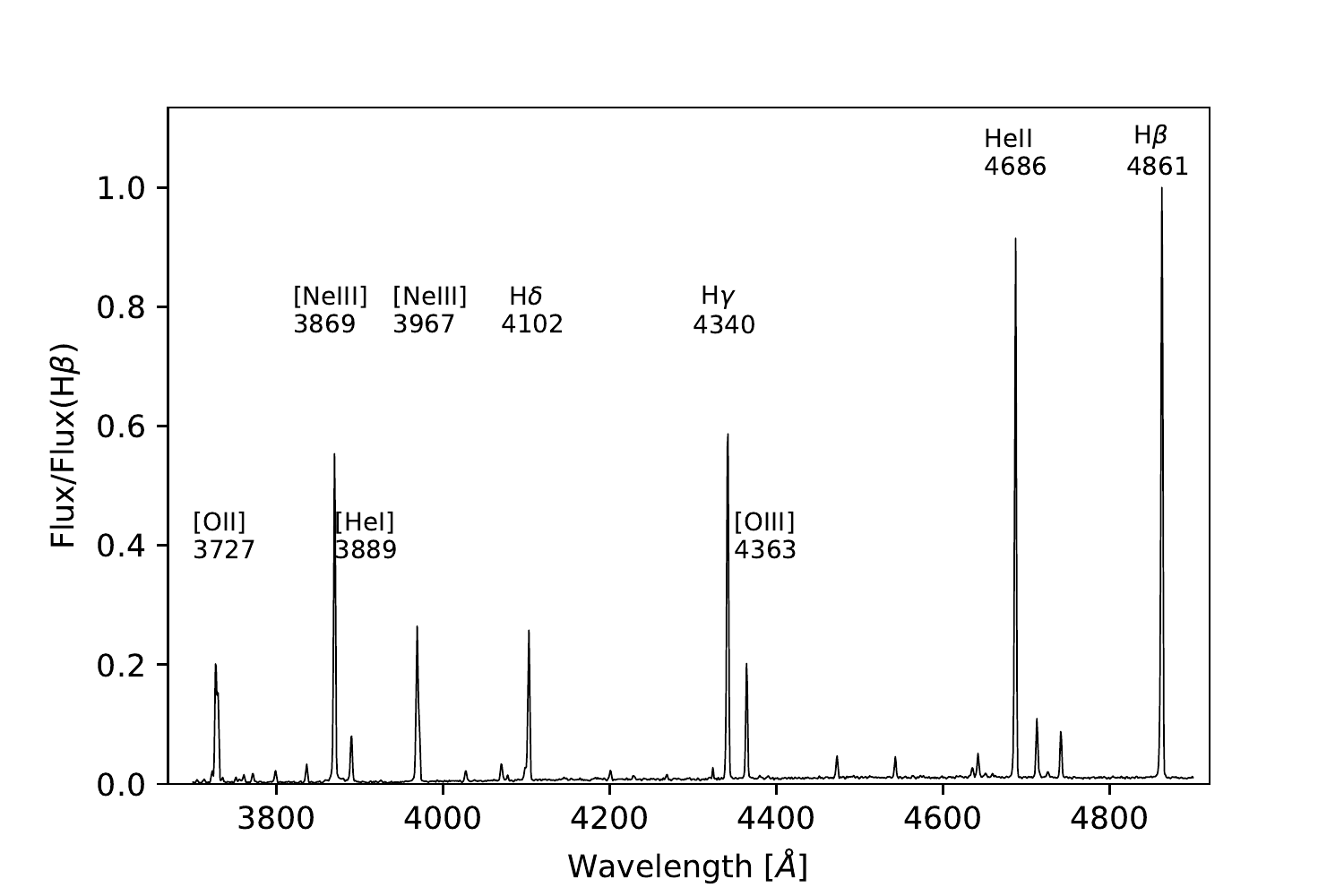}
                \caption{Reduced spectrum of the PN NCG\,2452. The flux was normalized to the height of the H$_\beta$ line.}
                \label{Fig:SpectrumPN}
    \end{figure}

%
\section{Measurements}
\label{s_meas}

        \subsection{NGC\,2453: radial velocities}
    \label{sub:RVClu}
  Radial velocities of program stars were measured using the Fourier cross-correlation technique \citep{Tonry79} via \textit{fxcor} IRAF task. The center of the correlation peak was fitted with a Gaussian profile. A grid of templates was prepared with synthetic spectra of solar metallicity drawn from the \citet{coelho2014}\footnote{http://specmodels.iag.usp.br/} library. The grid spanned the range from 375 to 500nm in step of 0.02~\AA, covering 3000~$\leq$~T$_\mathrm{eff}$~$\leq$~26000~K and 2.5~$\leq$~$\log g$~$\leq$~4.5, in steps of 2000~K and 0.5~dex, respectively. 
  Most of the targets were better cross-correlated with the template at $T_\mathrm{eff}$=22000~K, $\log g=4.5$, except for MSP\,211 and NGC\,2453~16, which required a cooler model (6000 and 10000~K, respectively), and the red giant TYC\,6548-790-1, for which the correlation height was maximized at $T_\mathrm{eff}=4000$~K and $\log g=2.5$. \citet{Moni11} and \citet{Morse91} showed that the exact choice of the template does not introduce relevant systematic error, although a mismatch between the target and the template spectral type can increase the resulting uncertainties.

The RV of hot stars was eventually measured with a CC restricted to the dominant Balmer lines (see MSP\,111 in Fig. \ref{Fig:spectra}), that is, in the intervals $4840-4885$~\AA\ ~(H$_{\beta}$), $4315-4365$~\AA\ ~(H$_{\gamma}$), $4075-4125$~\AA\ ~(H$_{\delta}$), and $3760-3995$~\AA\ ~(H$_{\epsilon}$ to H$_{12}$). The lines with hints of core emissions, namely H$_{\beta}$ and H$_{\gamma}$ in MSP\,74 were excluded from the CC. Spectral feature analyses for cold stars demanded more care. While they were bright stars, the low resolution blended the closest features (see TYC\,6548-790-1 spectrum in Fig. \ref{Fig:spectra}), although the stars were bright and the noise was not the dominant source of uncertainties in the optical range. Nevertheless, these stars were faint in the blue-UV edge of our spectra, where the camera was also less efficient (QE of 55\% at 3500~\AA~ against 80\% at 4000~\AA). In order to avoid possible sources of systematic error at the CCD borders, we measured the RVs using the wavelength interval $4000-4800$~\AA. The central peak of the CCF was higher than 0.95 for the target stars, indicating a high degree of similarity with the adopted template, except for TYC\,6548-790-1, for which it reached 0.82 only.

All RVs were measured relative to the solar system barycenter. Zero-point corrections were made using three standard stars of spectral types K and G (\citet{chubak2012}) treated in the same way as the cold stars described above. We found an average zero-point correction of $-9\pm2$~km~s$^{-1}$. The results are reported in Table~\ref{Tab:RV_Dis}. The final error was obtained as the quadratic sum of the most relevant sources of uncertainties, namely the measurement error obtained in the CC procedure, the zero-point correction uncertainty, and the wavelength calibration error (although this last resulted negligible).

Radial velocity measurements were performed on both targets and additional stars. However, the results for the latter are not reliable, because the random location of their PSF centroid in the spectrograph slit could easily have introduced a large systematic uncertainty on their RVs. In fact, the target stars MSP\,132 and MSP\,85 showed a very different RV when they fell as additional objects in other frames, and the two measurements of the additional star 2MASS\,J07473034-2711464 differ noticeably (see Table~\ref{Tab:RV_Dis}). Hence, we report the results for all measurements, but exclude the additional stars from the RV analysis.

%
        \begin{figure}
        \centering
                \includegraphics[trim = 10mm 0mm 0mm 10mm, clip,angle=0,width=10cm]{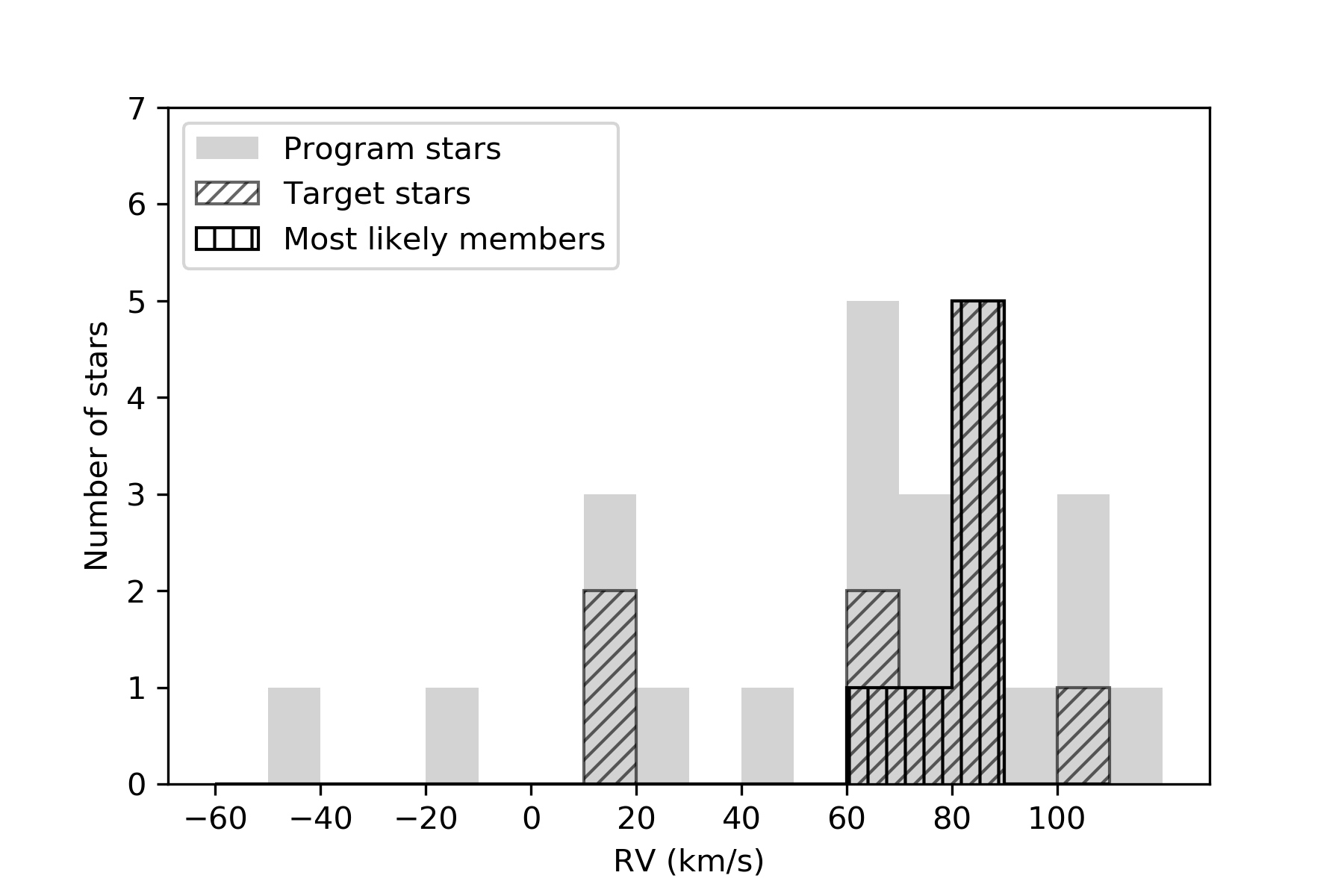}
        \caption{Radial velocity distribution for program and target stars of NGC~2453.}
        \label{Fig:Histo}
        \end{figure}

%
\subsection{\textsl{NGC\,2453: temperature and gravities}}
\label{sub:TgClu}

    The fundamental parameters (temperature, gravity, and rotation velocity) of the most likely cluster members (see section \ref{sec:Results}) were measured as in \citet{Moni17}, by means of the routines developed by \citet{Bergeron92} and \citet{Saffer94}, as modified by \citet{Napiwotzki99}. Briefly, the available Balmer and He lines were fitted simultaneously with a grid of synthetic spectra obtained from model atmospheres of solar metallicity, computed with ATLAS9 \citep{Kurucz93}. 
    The stellar rotation projected along the line of sight, $v\sin i$, is not a fit parameter but an input quantity of the routines. It was therefore varied manually until finding the value which returned the solution with the lowest $\chi^2$. 
    The results are given in Table~\ref{tab:Temperatures}, along with the photometric data of the targets from our optical photometry. The algorithm does not take into account possible sources of systematic error, such as the flat fielding procedure, the continuum definition, and the spectrum normalization.  Hence, the errors returned by the routine were multiplied by a factor of three to derive a more realistic estimate of the uncertainties (see, e.g.,  \citep{Moni17}). 

    The stellar temperature is mainly derived from the relative intensity of the Balmer lines, which is well measured in our spectra. On the contrary, surface gravity is estimated from the width of these features, but the method was insufficient to properly resolve its effects. In fact, we found a general underestimate of $\log g$ by about 0.2~dex when compared to expectations for MS objects ($\log g\approx$4.2), possibly due to the combination of a low spectral resolution and unresolved effects of stellar rotation. However, \citet{Zhang17} suggested that the method might be underestimating the surface gravity of MS stars by $\sim$0.1~dex even at very high spectral resolution.

%
\subsection{\textsl{NGC\,2452: radial velocity}}
\label{sub:RVPla}

The spectrum of NGC\,2452 is shown in Fig.~\ref{Fig:SpectrumPN}. Bright emission lines of [OII] (3727~\AA), [NeIII] (3967~\AA, 3869~\AA), HeII (4686~\AA), and the Balmer lines   H$_\beta$ (4861~\AA), H$_\gamma$ (4340~\AA) and H$_\delta$ (4102~\AA) can be easily identified. For a more detailed description of NGC\,2452 spectra in different locations we refer the reader to Table IV in \citet{Aller1979}.

    The RV of the PN was measured by CC with a synthetic spectrum. This was built adding up Gaussian curves with widths and heights equal to the observed features, but centered at the laboratory wavelengths taken from NIST Atomic Spectra Database Lines Form\footnote{https://www.nist.gov/pml/atomic-spectra-database}. 
    The reduction returned five spectra for NGC\,2452, namely a wide and narrow extraction for both the northern and the central regions, and a narrow extraction for the southern one. The measurements were repeated independently for the five spectra, to verify if the results could be affected by the internal kinematics of the nebula. We did not detect any systematic error between the spectra beyond fluctuations compatible with observational errors. The final estimate was obtained from the average of these measurements, and is reported in Table~\ref{table:RVNebula} along with previous values from the literature. Our final result is RV=62$\pm$2~km~s$^{-1}$, in good agreement with the weighted mean of literature results of 65$\pm$2~km~s$^{-1}$.

%
        \begin{figure}
        \centering
                \includegraphics[trim = 0mm 35mm 0mm 0mm, clip, angle=-90,width=10cm]{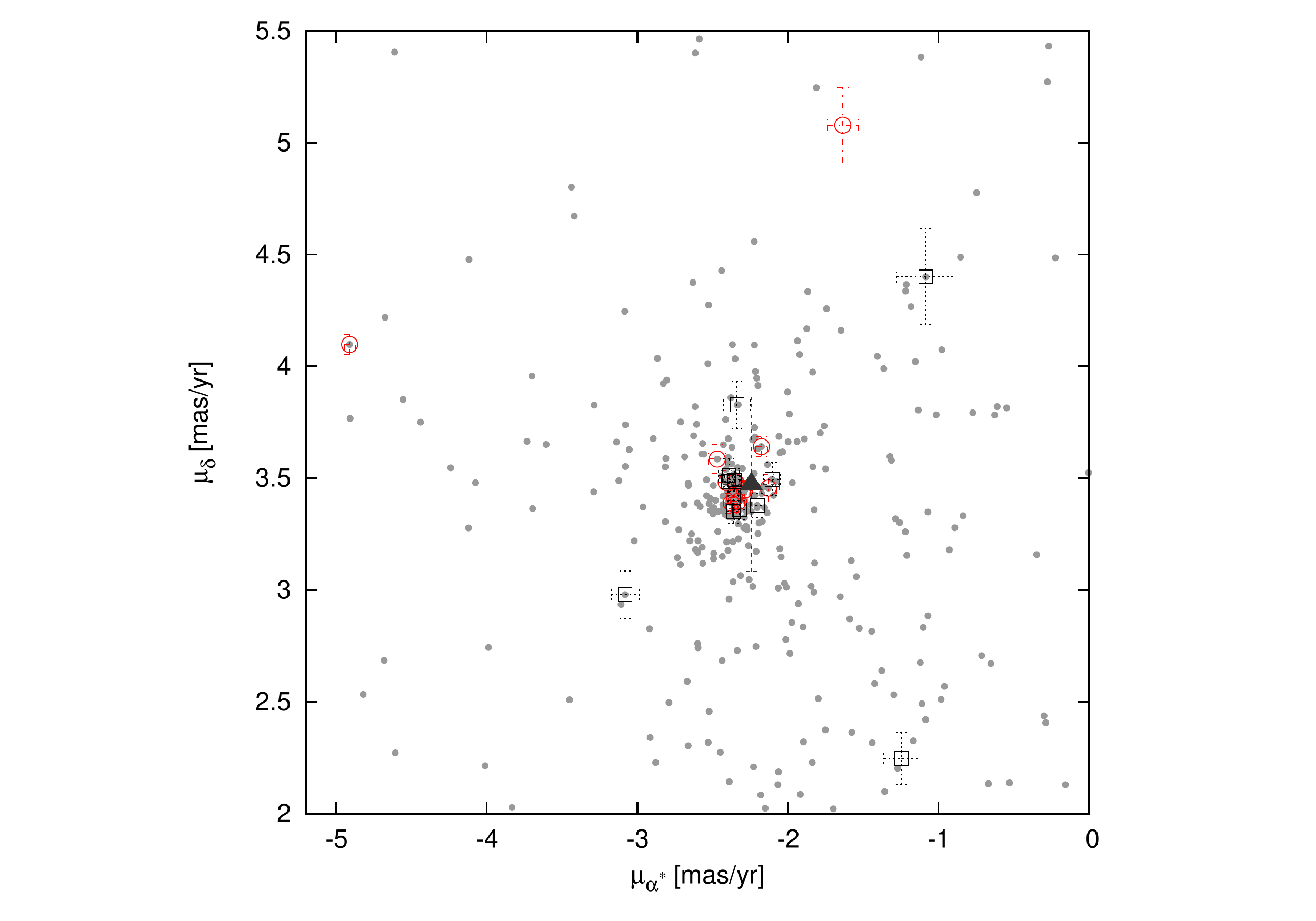}
                \caption{Proper motion of stars within $2\farcm5$ of the NGC\,2453 center (gray points), from the \textit{Gaia} DR2 catalog. The open red circles and black squares show the position of the target and additional stars, respectively. The triangle indicates the MF54 star.
        \label{Fig:ProperMotion}}
        \end{figure}

%
        \begin{table*}
        \tiny
                \centering
                \caption{Photometric data, radial velocities, and distances 
                        of the program objects.}
                \label{Tab:RV_Dis}
        \begin{spacing}{1.7}
                \begin{tabular}{@{}lcccccccl@{}}
                \hline\hline
                        Name & Type & \begin{tabular}[c]{@{}c@{}}J\\ (mag)\end{tabular} & \begin{tabular}[c]{@{}c@{}}J-H\\ (mag)\end{tabular} & \begin{tabular}[c]{@{}c@{}} $\mu^{\dagger}_{\alpha^{*}}$ \\ (mas yr$^{-1}$)\end{tabular} & \begin{tabular}[c]{@{}c@{}} $\mu^{\dagger}_{\delta}$\\ (mas yr$^{-1}$)\end{tabular}  & \begin{tabular}[c]{@{}c@{}}RV\\ (km/s)\end{tabular} & \begin{tabular}[c]{@{}c@{}}Distance\\ (kpc)\end{tabular} & \multicolumn{1}{c}{Note$^{\ddagger}$} \\ \hline
                         & \multicolumn{1}{l}{} &  &  &  &  &  \\
                        TYC\,6548-790-1  & T & 6.73  $\pm$ 0.02 & 0.85 $\pm$ 0.06 & $-2.33 \pm 0.04$ &       3.40 $\pm$      0.05 & 80 $\pm$ 10 & 5.2 $^{6.2}_{4.4}$ & MLM \\
                        MSP\,111                 & T & 11.81 $\pm$ 0.02 & 0.10 $\pm$ 0.05 & $-2.35 \pm 0.04$ &    3.41 $\pm$      0.05 & 69 $\pm$ 4 & 5.4 $^{6.5}_{4.6}$ &  MLM\\
                        MSP\,112                 & T & 12.28 $\pm$ 0.03 & 0.12 $\pm$ 0.06 & $-2.38 \pm 0.04$ &    3.39 $\pm$      0.04 & 89 $\pm$ 6 & 4.6 $^{5.4}_{4.0}$ &  MLM\\
                        MSP\,126                 & T & 12.28 $\pm$ 0.02 & 0.09 $\pm$ 0.05 & $-2.18 \pm 0.04$ &    3.64 $\pm$      0.04 & 89 $\pm$ 7 & 4.2 $^{4.7}_{3.7}$ &  MLM\\
                        MSP\,159                 & T & 12.13 $\pm$ 0.03 & 0.09 $\pm$ 0.06 & $-2.41 \pm 0.05$ &    3.48 $\pm$      0.05 & 88 $\pm$ 8 & 4.4 $^{5.3}_{3.8}$ &  MLM\\
                        MSP\,85                  & T & 12.53 $\pm$ 0.02 & 0.08 $\pm$ 0.05 & $-2.29 \pm 0.04$ &    3.45 $\pm$      0.04 & 87 $\pm$ 7 & 4.6 $^{5.3}_{4.1}$ &  MLM\\
                                         & A &                                  &                                   &                              &                                       & 117$\pm$ 7 &                             &  \\
            MSP\,132             & T & 11.83 $\pm$ 0.02 & 0.10 $\pm$ 0.05 & $-2.13 \pm 0.06$ &    3.46 $\pm$      0.06 & 72 $\pm$ 4 & 4.7 $^{5.8}_{3.9}$ &   MLM\\
                                                         & A &                                  &                                   &                              &                                       & 28 $\pm$ 4 &                     &   \\
            NGC\,2453~55         & T & 12.82 $\pm$ 0.05 & 0.15 $\pm$ 0.10 & $-1.64 \pm 0.10$ &    5.08 $\pm$      0.20 & 64 $\pm$ 6 & 11.0 $^{14.4}_{8.4}$ & NM, $\varpi$ < 0\\
            MSP\,57             & T & 11.71 $\pm$ 0.02 & 0.09 $\pm$ 0.05 & $-2.47\pm 0.06$ &     3.59 $\pm$      0.06 & 103$\pm$ 5 & 1.2 $^{1.2}_{1.1}$ & NM \\
                        NGC\,2453~16    & T & 12.11 $\pm$ 0.04 & 0.15 $\pm$ 0.09 & $-4.91 \pm 0.04$ &       4.10 $\pm$      0.05 & 16 $\pm$ 2 & 1.3 $^{1.3}_{1.2}$ & NM \\
            MSP\,211            & T & 12.68 $\pm$ 0.03 & 0.15 $\pm$ 0.06 & $-2.35 \pm 0.03$ &    3.48 $\pm$      0.04 & 18 $\pm$ 8 & 4.4 $^{4.8}_{4.0}$ & NM \\
                        2MASS\,J07473821-2710479 & A & 15.23 $\pm$ 0.06 & 0.25 $\pm$ 0.10 & $-2.39 \pm 0.07$ &    3.51 $\pm$      0.08  & 72 $\pm$  6 & 2.8 $^{3.4}_{2.4}$ &  \\
                        2MASS\,J07473390-2710060 & A & 15.36 $\pm$ 0.05 & 0.37 $\pm$ 0.10 & $-3.08 \pm 0.09$ &    2.98 $\pm$      0.10  & 66 $\pm$ 15  & 3.4 $^{4.5}_{2.7}$ &  \\
                        MSP\,52                 & A & 14.24 $\pm$ 0.08 & 0.17 $\pm$ 0.20 & $-2.36 \pm 0.04$ &    3.35 $\pm$      0.05 & $-11 \pm 4$  & 4.2 $^{4.9}_{3.7}$ &  \\
                        MSP\,272                & A & 12.89 $\pm$ 0.03 & 0.25 $\pm$ 0.05 & $-1.08 \pm 0.20$ &    4.40 $\pm$      0.20 & $-50 \pm 9$  & 0.9 $^{1.0}_{0.8}$ &  \\
                        MSP\,76                 & A & 12.91 $\pm$ 0.02 & 0.16 $\pm$ 0.04 & $-2.40 \pm 0.03$ &    3.48 $\pm$      0.03 &  18 $\pm$ 3  & 4.1 $^{4.4}_{3.7}$ &  \\
                        MSP\,141                & A & ---                          & ---                          & $-2.36 \pm 0.04$ &   3.49 $\pm$      0.05 &  44 $\pm$ 4  & 4.2 $^{4.9}_{3.8}$ &  \\
                        MSP\,74                 & A & 11.87 $\pm$ 0.03 & 0.21 $\pm$ 0.06 & $-2.36 \pm 0.03$ &    3.42 $\pm$      0.05 &  103$\pm$ 6  & 3.5 $^{3.9}_{3.2}$ &  \\
                        2MASS\,J07473034-2711464 & A & 14.68 $\pm$ 0.03 & 0.05 $\pm$ 0.05 & $-2.34 \pm 0.09$ &    3.83 $\pm$ 0.10  & 97 $\pm$ 5 & 2.0$^{2.3}_{1.8}$ &  \\
                                                    & A &                                  &                              &                                       &                                         & 70 $\pm$ 4 &                                    &  \\
            2MASS\,J07473176-2710057 & A & 14.58 $\pm$ 0.07 & 0.36 $\pm$ 0.20 & $-2.11 \pm 0.06$ &       3.50 $\pm$ 0.07  & 66 $\pm$ 6  & 3.6$^{4.3}_{3.1}$ &  \\
                        MSP\,204                & A & 14.18 $\pm$ 0.07 & 0.24 $\pm$ 0.20 & $-2.20 \pm 0.05$ &    3.38 $\pm$ 0.05 & 101 $\pm$ 5  & 4.2$^{4.9}_{3.7}$ &  \\
                        MSP\,223                & A & 14.06 $\pm$ 0.04 & 0.16 $\pm$ 0.08 & $-2.32 \pm 0.04$ &    3.36 $\pm$ 0.04 & 64 $\pm$ 4  & 3.7$^{4.1}_{3.3}$ &  \\ \hline
           
            MF54                        & - & 10.44 $\pm$ 0.03 & 0.17 $\pm$ 0.06 & $-2.24 \pm 0.20$ &       3.47 $\pm$ 0.40 & 67 $\pm$ 14$^{\dagger\dagger}$ &  4.2$^{6.5}_{2.9}$  &  $\varpi / \sigma_{\omega}$=0.96 \\ \hline \hline
                \end{tabular}
                \end{spacing}
        \raggedright{$^{\dagger}$ Data from \textit{Gaia} DR2. \\
                                 $^{\ddagger}$ MLM: Most Likely Member; NM: Non Member.\\
                     $^{\dagger\dagger}$ Data from \cite{moffat1974}
                     }

        \end{table*}

%
\begin{table*}
\small
\centering
        \caption{Derived parameters of the most likely members stars.}
        \label{tab:Temperatures}
        \begin{tabular}{@{}lcccccc@{}}
        \hline
        \multicolumn{1}{c}{Star} & $V$ & $(B-V)$ & $(U-B)$ & $T_\mathrm{eff}$   & $\log g$                & $v\cdot\sin i$  \\ 
        \multicolumn{1}{c}{}     &     &         &         & K                  & dex            & km~s$^{-1}$                            \\ \hline
TYC 6548-790-1           & 10.47   & 2.08    &  $1.73$   & ---                & ---                      &---                             \\
MSP85                    & 13.15   & 0.24    & $-0.40$   & 17700 $\pm$ 200    & 3.92 $\pm$ 0.03    &30                                      \\
MSP111                   & 12.66   & 0.31    & $-0.34$   & 16700 $\pm$ 300    & 3.63 $\pm$ 0.06    &90                                      \\
MSP112                   & 13.09   & 0.30    & $-0.33$   & 16600 $\pm$ 300    & 3.79 $\pm$ 0.06    &150                             \\
MSP126                   & 12.99   & 0.25    & $-0.40$   & 17800 $\pm$ 300    & 3.95 $\pm$ 0.06    &20                                      \\
MSP132                   & 12.51   & 0.26    & $-0.40$   & 16600 $\pm$ 200    & 3.90 $\pm$ 0.03    &160                             \\
MSP159                   & 12.79   & 0.24    & $-0.41$   & 17700 $\pm$ 300    & 3.86 $\pm$ 0.06    &40                                      \\ \hline
        \end{tabular}
\end{table*}

%
\section{Results}
\label{sec:Results}
The RV distribution of our program stars is shown in Fig.~\ref{Fig:Histo}, while the proper motions drawn from the \textit{Gaia} DR2 catalog are plotted in Fig.~\ref{Fig:ProperMotion}. Almost half of the RVs are comprised between 60 and 90~km~s$^{-1}$, where previous estimates of the cluster RV are found \citepalias{moffat1974,bidin2014}, while most of the program stars in the proper motion diagram cluster around ($\mu _\alpha \cos{\delta},\mu_\delta)\approx(3.5,-2.5)$~mas~yr$^{-1}$. The distances derived from \textit{Gaia} parallaxes are also listed in Table~\ref{Tab:RV_Dis}, and they are in the range 4.2-5.4~kpc for most of the targets.

The very high RV (103$\pm$5~km\,s$^{-1}$) and small distance (1.2$^{1.2}_{1.1}$~kpc) of the star MSP\,57 indicate that this is probably not a cluster member. The targets NGC\,2453~16 and MSP~211 are also suspected to be field stars due to their low RV (RV=16$\pm$2 and 18$\pm$8~km~s$^{-1}$, respectively), and for the former this conclusion is reinforced even by a discrepant distance and proper motion. In addition, NGC\,2453~55 lies far from the bulk of our sample in the proper motion plot, although its RV is compatible with it, and its uncertain distance does not provide additional information. These four stars were therefore labeled as ``non-member'' (NM) in Table~\ref{Tab:RV_Dis}, and excluded from further analysis. We are thus left with seven stars whose RVs, distances, and proper motions are very consistent, and these are considered ``Most Likely Members'' (MLM). Their RV distribution is shown with a vertically striped area in Fig. \ref{Fig:Histo}.

The RVs of stars in the field of NGC\,2453 were previously measured by \citetalias{bidin2014} using CCF from the H$_{\alpha}$ line. The authors estimated RV=73$\pm$5 and 66$\pm$8~km~s$^{-1}$ for TYC\,6548-790-1 and MSP\,111, respectively, in agreement with this work despite the large uncertainties. On the other hand, their result for MSP\,57 (RV=70$\pm$9~km~s$^{-1}$) disagrees with ours. The authors considered this star as a probable cluster member, but new data from \textit{Gaia} DR2 locate it at about 1.2~kpc, too close for an association with the cluster, and its membership is not supported. On the other hand, \citetalias{bidin2014} classified the star MSP\,159 as a nonmember, because its proper motion from the PPMXL catalog \citep{roeser2010} was clearly offset from the bulk of their sample. However, the accurate measurements from the \textit{Gaia} DR2 catalog indicate a proper motion consistent with MLM stars, along with compatible RV and distance. Regarding the red giant star TYC\,6548-790-1, \citet{mermilliod2001} and \textit{Gaia} DR2 obtained RVs of $85.2\pm0.3$~km~s$^{-1}$ and $85.5\pm0.3$~km~s$^{-1}$, respectively, in good agreement with ours. We added the star NGC\,2453\,54 (hereafter MF\,54) to our sample both in Table~\ref{Tab:RV_Dis} and Fig.~\ref{Fig:ProperMotion}, although its RV was measured by \citet{moffat1974} but not by us.  We return to this object in Sect.
\ref{sec:discussion}.

Finally, the RV of NGC\,2453 was computed using target stars labeled as MLM. We found a weighted mean of RV=$78\pm3~km~s^{-1}$, where the uncertainty is the statistical error on the mean. Table~\ref{table:RVNebula} compares our result with those available in the literature and reveals that our estimate differs from previous ones. These latter however were obtained from only one or two stars, whose cluster membership was inevitably uncertain. Our result, on the contrary, is based on a sample of seven stars with consistent RVs, proper motions, and parallax-based distances.

From the \textit{Gaia} measurements for our program stars, the cluster distance and proper motion can also be estimated. Despite the large errors on distances, the modal values of all MLM stars are close each other and they differ less than their respective uncertainties, suggesting that the latter could have been overestimated. We adopted the weighted means of MLM stars and the respective errors-on-the-mean as best estimates of the cluster value and their uncertainties, respectively, obtaining $d=4.7\pm0.2$~kpc, $\mu_{\alpha^{*}}=-2.30\pm0.04$~mas~yr$^{-1}$, and $\mu_{\delta}=3.47\pm0.03$~mas~yr$^{-1}$.

%
\begin{table}
\small
\centering
\caption{Literature results for the RV of the PN NGC\,2452 and the OC NGC\,2453.}
\label{table:RVNebula}
\begin{tabular}{@{}llc@{}}

\textbf{PN NGC~2452}                      &                  & \\ \hline

Literature                                & RV (km s$^{-1}$) & \\ \hline

\citet{wilson1953}                & 68.0 $\pm$ 2.5   & \\
\citet{meatheringham1988}             & 62.0 $\pm$ 2.8   & \\
\citet{durand1998}                    & 65 $\pm$ 3       & \\
                                  &                  & \\
\multicolumn{1}{r}{Literature Average}& 65 $\pm$ 2      & \\
\multicolumn{1}{r}{This Work}           & 62 $\pm$ 2 \\ \hline
\\
\\
\textbf{OC NGC~2453}    &                       &  \\ \hline
Literature                              & RV (km s$^{-1}$)      &       Number of stars \\ \hline
\citet{moffat1974}      & 67 $\pm$ 14           &       1         \\
\citet{bidin2014}       & 68 $\pm$ 4            &       2         \\

                        &                       &                 \\

\multicolumn{1}{r}{This Work} & 78 $\pm$ 3      &       7       \\\hline

\end{tabular}
\end{table}

%
        \begin{figure*}
        \centering
                \includegraphics[width=18cm]{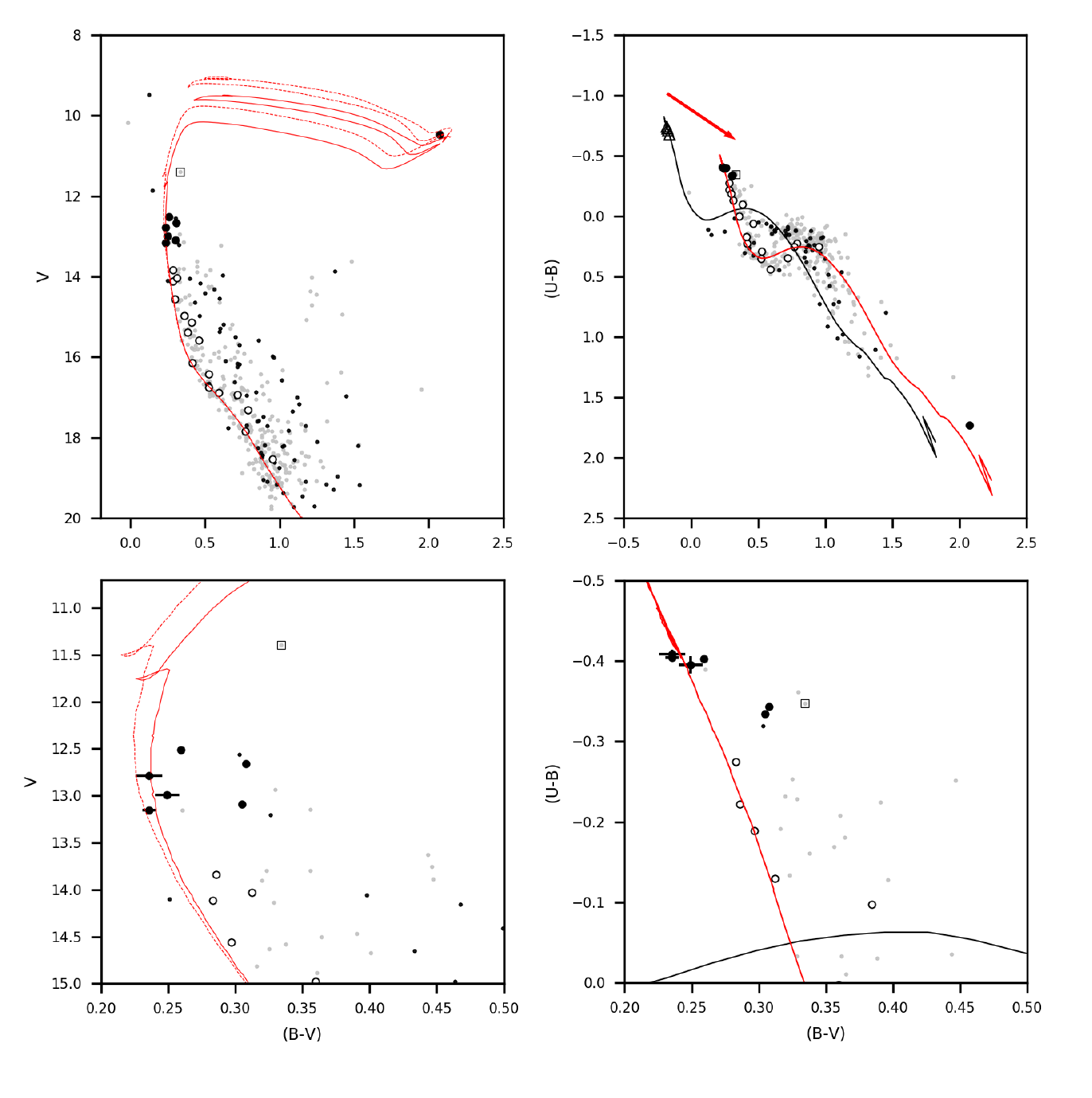}
                \caption{CMDs and TCDs of NGC\,2453. {\it Left panels}: The $V$-$(B-V)$ CMD. Dashed and solid lines depict isochrones of 40~Myr and 50~Myr, respectively, shifted in magnitude for a distance of 4.7~kpc. {\it Right panels}: $(U-B)$-$(B-V)$ TCDs. Black and red lines depict intrinsic and reddened isochrones, respectively, and the arrow shows the reddening direction. {\it Bottom panels}: Zoomed region of the upper panels around the MLM stars. Light gray dots indicate the stars in the field along the line of sight of the cluster, black filled circles show the MLM stars, and open circles indicate stars with proper motions within 2$\sigma$ of the cluster. Black empty dots are stars with upper distance errors $\leqslant$3.5~kpc from \textit{Gaia} DR2, and the star in the square is MF54. PARSEC + COLIBRI isochrones from \citet{marigo2017} have been fitted to MLM stars.}
                \label{Fig:CDMCCDISO}
    \end{figure*}

%
\subsection{Fundamental parameters}
\label{sub:Parameters}

NGC\,2453 has a great record of observations, but its fundamental parameters have proven difficult to establish, in part because of the complex mix of stars at different distances and reddening lying along the line of sight. 

In this work, we overcame the problems of field contamination estimating the cluster distance from the parallax-based \textit{Gaia} distances of spectroscopically confirmed members. With this information, we can thus determine the age and reddening of the system from isochrone fitting of our $UBVRI$ photometry, relying again on the constraints provided by the \textit{Gaia} database and our spectroscopic results. PARSEC + COLIBRI isochrones \citep{marigo2017} were used in this process.

The upper panels of Fig.~\ref{Fig:CDMCCDISO} show the $V-(B-V)$ CMD and the $(U-B)-(B-V)$ TCD of the cluster area. MLM stars have been depicted as black circles. The TCD (top-right panel) reveals the presence of at least two groups of stars with very different reddening. To identify the cluster sequence, we selected stars with \textit{Gaia} proper motion within 2$\sigma$ of the cluster value (identified as the mean of the MLM stars in Table~\ref{Tab:RV_Dis}), with proper motion error lower than 0.1~mas~yr$^{-1}$, and \textit{Gaia} distance close to $d=4.7\pm0.2$~kpc. These stars are depicted in Fig.~\ref{Fig:CDMCCDISO} as open circles. To identify foreground stars, we also selected those whose distance confidence interval had an 
upper edge (upper index in Table 1) lower than 3.5~kpc, and we indicated them with black dots in the diagrams. Indeed, most of these stars are better described by a less reddened sequence than the bona-fide cluster members (open circles), although a few field stars might still be contaminating the latter sample. The brighter MLM stars and the additional open circles thus identify the cluster loci in the TCD.

The intrinsic theoretical isochrone is shown in the TCD of Fig. \ref{Fig:CDMCCDISO} as a black solid curve, while the red one indicates the same model after applying the final reddening solution. The triangles on the intrinsic isochrone correspond to the points at the same temperature range as our spectroscopic estimates for MLM stars (see Table \ref{tab:Temperatures}), that is, $\log(T_{\mathrm{eff}})=[4.23,4.25]$.

We determined the color excesses $E_{U-B}$ and $E_{B-V}$ from the difference of the average color index for MLM stars (black circles), and for isochrone points at the same temperature (black triangles). We thus derived the slope $E_{U-B}/E_{B-V}$ of the reddening vector in the TCD. The bottom-right panel of Fig.~\ref{Fig:CDMCCDISO} shows a zoomed region of the TCD, focused on the MLM stars, where it appears clear that three MLM stars (namely MSP\,111, MSP\,112 and MSP\,132) are found at redder colors than the others, possibly due to stellar rotation effects \citep{bastian2009} or the presence of a cooler companion \citep{yang2011}. Table \ref{tab:Temperatures} shows that these stars as indeed fast rotators. As a consequence, only the slow-rotating MLM stars were used in the process. We obtained a slope of $E_{U-B}/E_{B-V}=0.78\pm0.09$, with $E_{B-V}=0.42\pm0.01$. This result agrees well with \citet{turner2012}, who established localized reddening laws described by E$_{U-B}$/E$_{B-V}$ = 0.77 and $R_V$ = 2.9 for the third galactic quadrant (\citealt{turner2014testing}; \citealt{carraro2015}), which is adopted here. The resulting extinction is $A_V=1.22\pm0.03$~mag. This result, together with the distance derived in this work, fits the general Galactic extinction pattern determined by \citet{neckel1980} very well, even though the authors did not study the NGC\,2453 region (l=343\degr, b=$-1$\degr). According to their work, the Galactic region near the cluster line-of-sight (l=342\degr, b=0\degr) has an extinction $A_V\approx 1$ up to $\sim$5~kpc, and it increases at a further distance to $A_V\approx 2$ at about 6~kpc and beyond. In contrast, the next region closest to the cluster area (l = 345\degr, b = 0\degr) shows an extinction $A_V\approx 1.5$ between 2 and 6~kpc, with slight variations at both $\sim$3.5 and $\sim$5.0~kpc. These results seem to be confirmed using the 3D map of interstellar dust reddening\footnote{http://argonaut.skymaps.info/} describe by \citet{Green2018}. The map shows a distance of $d=5.0$~kpc for a reddening of E$_{B-V}$ = $0.42 \pm 0.03$ in the same line of view of the cluster, in great agreement with our results.

Eventually, with the distance and reddening found so far, we fitted slow rotator MLM and bona-fide cluster stars in the CMD, with age as the only free parameter. We find that an age in the range $\tau\approx 40-50$~Myr is the best solution, which accurately reproduces the observed sequence of stars (see left panel of Fig.~\ref{Fig:CDMCCDISO}).

%
\begin{table}
\small
\centering
\caption{Parameters estimated for NGC 2453}
\label{table:Literature}
\begin{tabular}{llcl}
\hline
Reference               & \multicolumn{1}{c}{$E_{(B-V)}$} & \multicolumn{1}{c}{$\tau$ ~(Myr)} & \multicolumn{1}{c}{$d$ (kpc)}  \\
\hline
\\
\cite{seggewiss1971}           & 0.48                     & --          & 1.5      \\
\cite{moffat1974}       & 0.47 $\pm$ 0.04 & 40          & 2.9 $\pm$ 0.5      \\
\cite{gathier1986}      & 0.49 $\pm$ 0.01 &  --         & 5.0 $\pm$ 0.6  \\
\cite{mallik1995}               & 0.47                    & 25          & 5.9 $\pm$ 0.5  \\
\cite{moitinho2006}             & --                      & 40          & 5.25                     \\
\cite{hasan2008}            & 0.47                        & 200         & 3.3                  \\
                                                &                                 &                      &                                \\
This Work               & 0.42 $\pm$ 0.01 & 40-50       & 4.7 $\pm$ 0.2          \\
\hline
\end{tabular}
\end{table}


\section{Discussion}
\label{sec:discussion}

\subsection{Cluster parameters}
\label{sub:isofit}

Our estimates of reddening, distance, and age for NGC\,2453 are compared with literature results in Table~\ref{table:Literature}. All previous studies were purely photometric, while we joined information from optical spectroscopy, $UBV$ photometry, and recent data from the \textit{Gaia} mission.

The distance and age derived here are roughly compatible with those found by \citet[][5.23~kpc and 40~Myr]{moitinho2006}, but the former is closer to the result of \citet[][$d=5.0\pm0.6$~kpc]{gathier1986}. However, the reddening derived by Gathier et al. (and in general, all estimates in the literature) is $\sim$15\% larger than ours. These authors based their results on five stars previously classified as cluster members by \citetalias{moffat1974}, namely NGC\,2453~7, 8, 28, 30 and 45 \citep{gathier1985}. However, \textit{Gaia} distances for the stars 28 and 30 ($1.1^{1.2}_{1,1}$~kpc and $7.8^{9.5}_{6.4}$~kpc, respectively) disagree with the estimates of Gathier et al. ($\sim$3.9 and 4.4~kpc, respectively), and they are much larger than the average value for our MLM stars. This suggests that some stars used in previous works to constrain the cluster parameters may not have been cluster members. Gathier found that the color excess $E_{B-V}$ of these two stars is the same ($\sim 0.51$), in spite of the huge distance discrepancy reported by \textit{Gaia}. On the other hand, \citet{mallik1995} showed that a reddening of $0.47$, as proposed by \citetalias{moffat1974}, produces reasonably good isochrone fits on the CMD. However, our analysis shows that such high values accurately fit the color of a group of stars that are displaced to redder colors than the rest of the MS, possibly due to their fast rotation or to the presence of a cool companion.

We indicated the evolved giant star MF54 observed by \citetalias{moffat1974} as an empty square in Fig. \ref{Fig:CDMCCDISO}, and as a black triangle in Fig.~\ref{Fig:ProperMotion}. These authors classified MF54 as a cluster member based on its spectral class (B5V:k) and a RV of $67 \pm 14$~km~s$^{-1}$. Its \textit{Gaia} DR2 proper motion and distance agree with the mean values obtained for the cluster (see Table~\ref{Tab:RV_Dis}), despite the large error bars. However, the fractional parallax error is extremely large ($\sim$ 118\%), and it contrasts with the typical errors for MLM stars ($\lesssim$ 25\%), which produce less reliable distance measurements \citep{BailerJones15}. Due to the high uncertainties in the measurements, the membership of MF54 is not completely clear, and therefore we did not take it into account during the isochrone fit procedure. Similarly, the red giant star TYC\,6548-790-1 was also excluded from the fit. This star could be variable (see \citetalias{bidin2014}), and as a consequence its photometric data may not be completely reliable. \citet{mallik1995} showed that the inclusion of one or both of these two stars during the isochrone fitting procedure can change the cluster age from 15 to 40~Myr.

\begin{figure}
        \centering
                \includegraphics[trim = 0mm 0mm 0mm 0mm,clip,angle=-90, width=9.5cm]{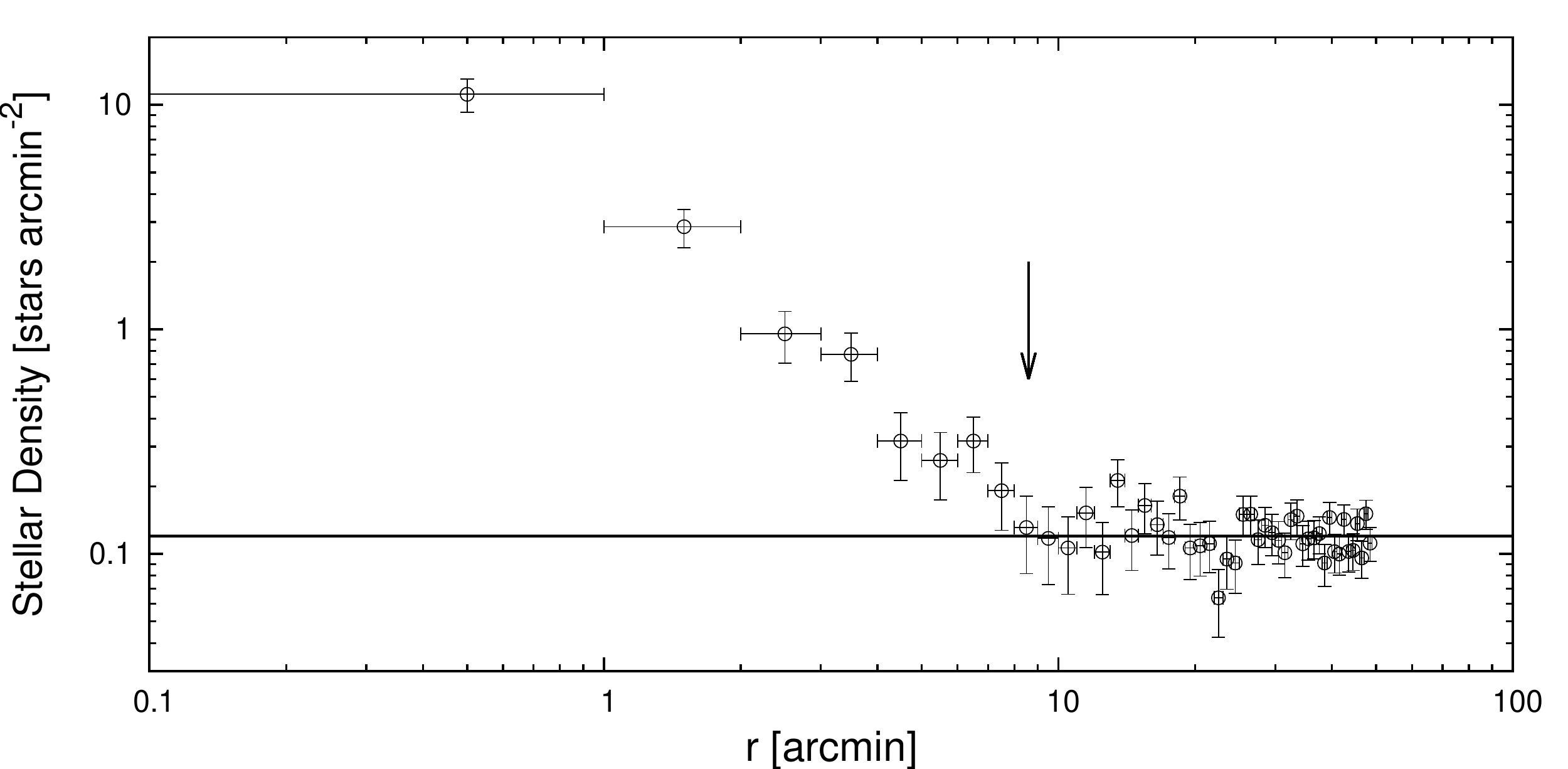}
                \caption{Radial density profile constructed for NGC\,2453 using proper motions from \textit{Gaia} DR2. The radial distance of NGC\,2452 is indicated with an arrow. The full line shows the field level as the average of all the points with r > 11$\arcmin$ .}
                \label{Fig:Density}
    \end{figure}

In Fig. \ref{Fig:Density} we analyze the radial density profile of the OC. Only stars with proper motion within 3$\sigma$ of the cluster value were selected. It is clear that the cluster population dominates  the  background up to approximately r$\sim 8\arcmin-10\farcm5$. The angular distance between PN NGC\,2452 and the center of the OC NGC\,2453 is $8\farcm5$, that is, within the coronal extent of the OC.

\subsection{Planetary nebula membership}
\label{sub:membership}
\citet{gathier1986} derived the reddening of the PN NGC\,2452 as $E_{B-V} = 0.43 \pm 0.05$, which is virtually the same found by us for the cluster. Nevertheless, the reddening-distance method used by \citet{gathier1986} for the PN leads to a distance of $d_{PN}=3.57\pm0.47$~kpc, which is confirmed with the more modern dust map by \citet{Green2018} ($d_{PN}=3.70$~kpc). Other authors adopted different methods, and found even smaller values \citep[see, e.g.,][]{acker1978,maciel1980,daub1982,stanghellini2008}.

Distance and proper motions from \textit{Gaia} DR2 to PN NGC\,2452 are not particularly reliable ($d_{PN}=2.4^{3.4}_{1.8}$~kpc,  $\mu_{\alpha}$=-2.5$\pm$0.2~mas~yr$^{-1}$ and $\mu_{\delta}$=3.5$\pm$0.2~mas~yr$^{-1}$).  Even though the central star for NGC\,2452 was a target of various photometric studies \citep[e.g.,][]{Ciardullo96,Silvotti96}, and its coordinates match  those from \textit{Gaia}  very well, \citet{kimeswenger2018} restrict the identification to PNe with photometric colors in the range $-0.65\leqslant (bp -rp)\leqslant -0.25$. Outside this interval, \textit{Gaia} DR2 cannot identify the central star correctly due to contamination of the H$_\alpha$+[NII] emission line of the PN envelope. The color index for NGC\,2452 is $(bp -rp)=0.07$, which is highly reddened. Therefore, any identification would most likely be incorrect. 

\begin{figure}
        \centering
                \includegraphics[trim = 0mm 0mm 0mm 10mm, width=10cm]{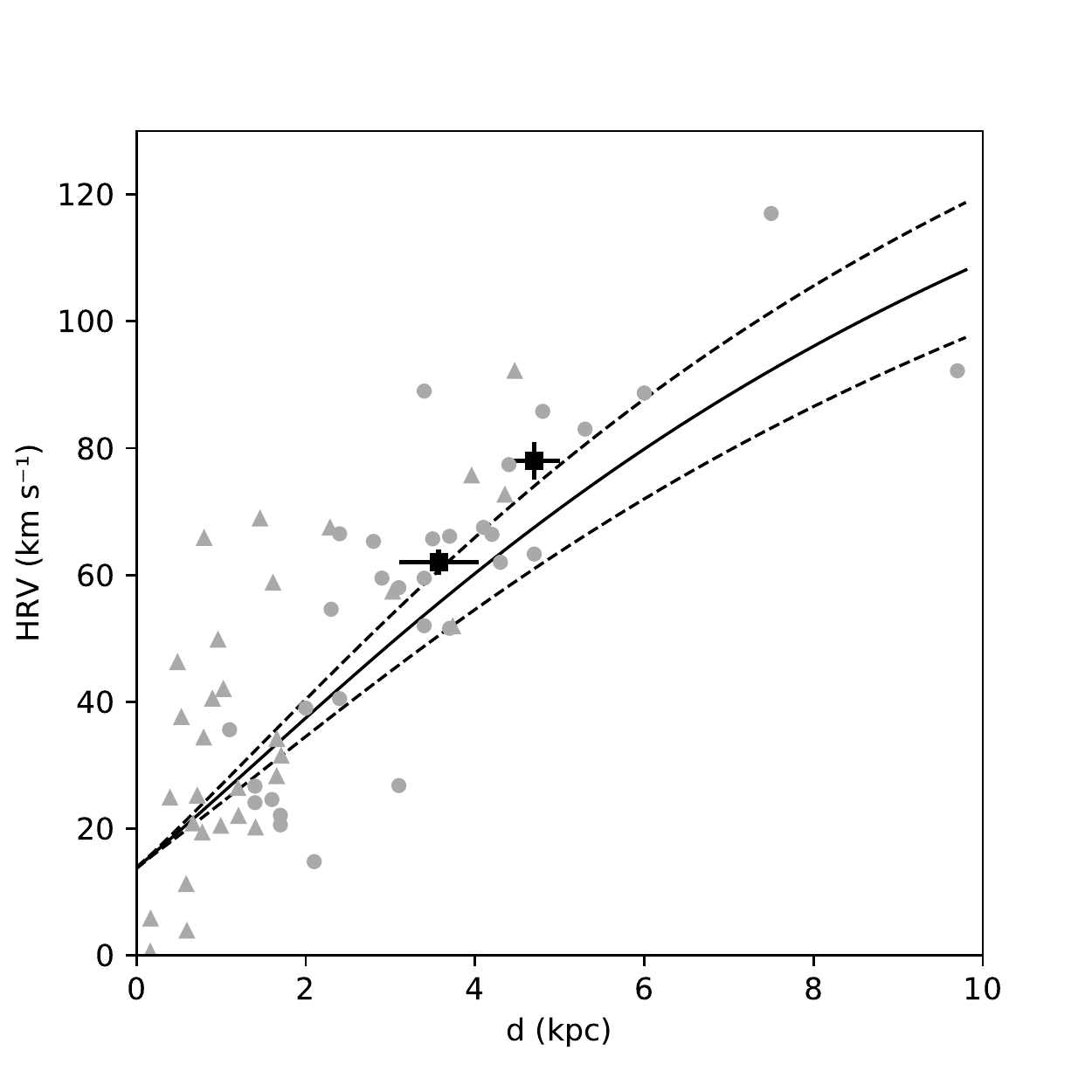}
                \caption{Distance--RV plot in the   direction of Puppis. The solid curve shows our theoretical model based on Galactic rotation, with the dashed curves used to indicate the $1\sigma$ propagation errors. Gray circles are classical Galactic Cepheids from \citet{mel2015} in the third quadrant with Galatic latitudes $-2\degree<b<2\degree$, while triangles are bright stars with available RVs from \textit{Gaia} DR2 with $242.5\degree<l<243.5\degree$ and $-1\degree<b<1\degree$. Squares with error bars show the position of NGC\,2452 and NGC\,2453.}
                \label{Fig:RV_D}
    \end{figure}

Figure~\ref{Fig:RV_D} shows that the RV of PN~NGC\,2452, along with the distance proposed by \citet{gathier1986}, closely match the distance--RV profile of the Galaxy arm in the Puppis direction. The profile was obtained assuming the rotation curve of \citet{brand1993}, the solar peculiar motion of \citet{schonrich2010}, $ R_\odot=8.0\pm0.3$~kpc, and $V_\mathrm{LSR}=220\pm20$~km~s$^{-1}$. In contrast, the cluster NGC\,2453 is consistent in both RV and distance computed here to be just behind NGC\,2452, and possibly a member of the Perseus arm, as  can be seen in Fig.~2 of \citet{moitinho2006}.

%

\section{Conclusions}
\label{sec:conclusions}

We present the results of distance analyses solving the longstanding discrepancy regarding the fundamental parameters of the OC NGC\,2453 and the debated cluster membership of the PN NGC\,2452, which were likely affected by the selection of cluster stars contaminated by field objects.

The study of RVs has often been required to confirm real PN/OC associations (see, e.g., \citealt{mallik1995}, \citealt{ majaess2007}, \citetalias{bidin2014}). When the RVs of the PN and the OC disagree, the membership is rejected (\citealt{kiss2008}, \citetalias{bidin2014}). 
The difference in RV between the PN (62$\pm$1~km~s$^{-1}$) and the cluster  (78$\pm$3~km~s$^{-1}$) is noticeable and highly significant ($\sim5\sigma$),  excluding a physical association between them.

All photometric diagrams show the presence of a robust group of foreground stars located at distances $\leqslant$3.5~kpc and contaminating the cluster field. According to the theoretical distance--velocity profile of the Galactic disk in the direction of Puppis, the RV we obtain for the PN NGC\,2452 is consistent with membership to this foreground population.

\begin{acknowledgements}
      This work has made use of data from the European Space Agency (ESA) mission \textit{Gaia} (\url{https://www.cosmos.esa.int/gaia}), processed by the \textit{Gaia} Data Processing and Analysis Consortium (DPAC, \url{https://www.cosmos.esa.int/web/gaia/dpac/consortium}); data from the Two Micron All Sky Survey, which is a joint project of the University of Massachusetts and the Infrared Processing and Analysis Center/California Institute of Technology, funded by the National Aeronautics and Space Administration and the National Science Foundation; and the SIMBAD and Vizier databases, operated at the CDS, Strasbourg, France. We thank the anonymous referee for his/her useful comments and suggestions. ESV thanks "Estrategia de Sostenibilidad, Universidad de Antioquia". DGD acknowledge support from the Plataforma de Movilidad Estudiantil y Acad\'emica of Alianza del Pac\'ifico and its chilean headquarters in the Agencia Chilena de Coperaci\'on Internacional para el Desarrollo - AGCID, as well as the support from call N$\degr$ 785 of 2017 of the colombian Departamento Administrativo de Ciencia, Tecnolog\'ia e Innovaci\'on, COLCIENCIAS. 
\end{acknowledgements}


%

\bibliographystyle{aa}
\bibliography{NGC2453}

\end{document}